# Topological Photonic Crystal of Large Valley Chern Numbers

Xiang Xi, Kang-Ping Ye, and Rui-Xin Wu*

School of Electronic Science and Engineering, Nanjing University, Nanjing 200023, China

* Corresponding author: rxwu@nju.edu.cn

**Abstract:** The recent realizations of topological valley phase in photonic crystal, an analog of gapped valleytronic materials in electronic system, are limited to the valley Chern number of one. In this letter, we present a new type of valley phase that can have large valley Chern number of two or three. The valley phase transitions between the different valley Chern numbers (from one to three) are realized by changing the configuration of the unit cell. We demonstrate that these new topological phases can guide the wave propagation robustly along the domain wall of sharp bent. Our results are promising for the exploration of new topological phenomena in photonic systems.

Photonic topological states, rooted from the studies of topological insulators in electronic systems, have opened up an intriguing way to control the motion of electromagnetic (EM) waves. The topological edge states resulting from the topological phases can route the wave propagation overcoming the backscattering and robust against defects [1-5], which attracts a great many interests. Up to now, many photonic topological phases have been proposed such as the quantum Hall (QH) phase [2-11], the quantum spin Hall (QSH) phase [12-19] and the quantum valley Hall (QVH) phase [20-32].

The symmetry plays a key role in designing the topological phases of photonic crystals (PCs). For example, in the honeycomb structured magnetic PCs, breaking the time-reversal (TR) symmetry will gap a pair of Dirac cones at high symmetry *K* and *K'* points, so that the Berry curvature has the same sign at these points and yield a non-zero topological invariant number, the Chern number, |*C*|=1. If the Dirac points are away from the high symmetry points, the Berry

curvature will have more extremes around the *K* and *K'* points, and the Chern number is greater than one [6, 7]. Similarly, the QVH phase is related to the breaking of inversion symmetry, which introduces a binary degree of freedom (DOF) in the PCs, an analog of the valley DOF spintronics in the electronic system. Valley labels the energetically degenerate yet inequivalent points in momentum space [20]. This new DOF also opens the Dirac cones at high symmetry *K* and *K'* points, but the Berry curvature at these two points has opposite sign, therefore the Chern number of the bandgap will be zero. However, the valley Chern number $C_v$ defined at the valley points is nonzero [22, 24]. The topological valley phase gets rid of the limitation of bias magnetic field, opening a path toward the topological phase in all-dielectric PC [20-32]. To date, the QVH phase in PCs is limited to $|C_v|=1$. Then, a natural question is that if the QVH phase can have a large valley Chern number.

In this work, we report a new type of valley phase that the valley Chern number can be $|C_v|=1, 2$ or 3 depending on the configuration of the unit cell. The variations of the valley Chern number are achieved by expanding or shrinking one set of rods of the hexamer. These new QVH phases are characterized by the Berry curvature in the first Brillouin zone, and further proved by the edge states at the domain wall according to the bulk-edge correspondence [33, 34]; the numbers of edge states are the same as the difference of valley Chern numbers across the domain wall. The robust wave transmissions of these QVH phases are demonstrated by the Z-shaped domain wall. Having band gaps with larger valley Chern number greatly expands the phases available for topological photonics.

Considering the PC structure is fabricated by the artificial molecules as shown in Fig. 1(a). Each molecule, composed of six neighboring rods, is a hexamer in the background air. The molecules are arranged in the hexagonal lattice, and the lattice constant is $a=10\sqrt{3}$ *mm*. Supposing the rods are made of YIG, whose relative permittivity and permeability are $\varepsilon_r = 15.26$ and $\mu_r=1$, respectively, at microwave frequencies. The radius of the rods is $r=0.2a$. The dimension of the molecule is measured by the distance between the rods' center and the center of the unit cell, denoted by *R* in the figure.

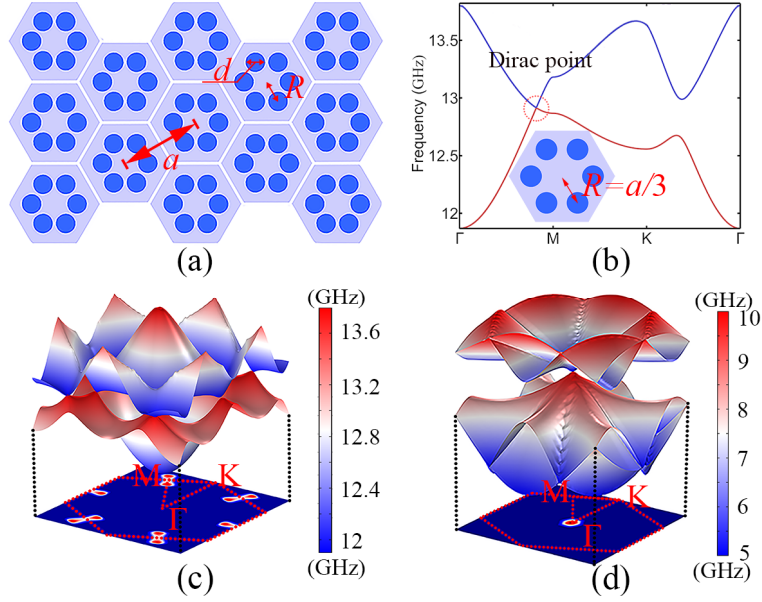

Fig. 1. (a) Schematic of 2D PC structure composed of hexamers of six ferrite rods is embedded in the air background. The white lines denote the edge of a unit cell. (b) Band structure of the PC at $R=a/3$. A Dirac point is away from the high symmetry points in the first Brillouin zone. (c) 3D band structure of the PC. Three pairs of Dirac points are between the two bands in the Brillouin zone. (d) 3D band structure of Ref. [13]. One pair of Dirac points presents at the $\Gamma$ point.

Figure 1(b) plots the band structure at $R = a/3$. A Dirac point presents at the frequency of 12.9 GHz in the band structure. This point is a general point in $k$-space, not at a high symmetry point as the works reported [20-32]. We note the arrangement of the hexamer in the unit cell is different from Ref. [13] where the hexamer has an additional rotation of $\pi/6$. Our configuration keeps primitive unit cell of six rods even when the Dirac point presents. Due to the symmetry of the system, three pairs of Dirac points emerge in the first Brillouin zone, as displayed in Fig. 1(c). In contrast, only one pair of Dirac points is in the configuration of Ref. [13] at the $\Gamma$ point as displayed in Fig. 1(d). The increment of the fold of degeneracy results in a large Chern number of the bands when they are gaped [6, 7].

To gap the Dirac points, one can break the time-reversal symmetry or the space inversion symmetry. After the symmetry breaking operation, the Dirac points is opened, and each degeneracy-lifting contributes a Berry flux of magnitude $\pi$ in each band [2-5], leading to a peak in the Berry curvature. Each peak contributes Chern number $|C|=1/2$, and when total Berry flux adds up to $2\pi$, the Chern number will be $|C|=1$ [8]. Here, we gap the Dirac points by breaking the inversion symmetry of the system. Shrinking or expanding the distance $R$ of the set of rods

(marked by red) and keeping the other set unchanged, the rotation symmetry of the system changes from original $C_6$ to $C_3$, and the inversion symmetry of the system is broken. This operation opens a gap at the Dirac point as displayed in Fig. 2(a), where the valleys appear at *K(K')* point. However, the valleys are different from those reported [20-32]. As an illustration, Fig. 2(b) plots the Berry curvature of the low band in the first Brillouin zone. We see one peak presents at the valley and the other three peaks around it. Because the system preserves time-reversal symmetry, the peaks up and down are in the same numbers, thus the Chern number of the band is zero. However, the band can be characterized by valley Chern number because the Berry curvature is distinguished at *K* and *K'* valleys. Fig. 2(b) shows near the *K* valley three peaks up and one peak down, while near the *K'* valley one peak up and three peaks down. Therefore, the valley Chern number of the band is $C_v = C_k - C_{k'} = 1-(-1) = 2$. Besides, simply rotating the original unit cell by 60° will reverse the valley Chern number, from $C_v=2$ to $C_v=-2$ [35].

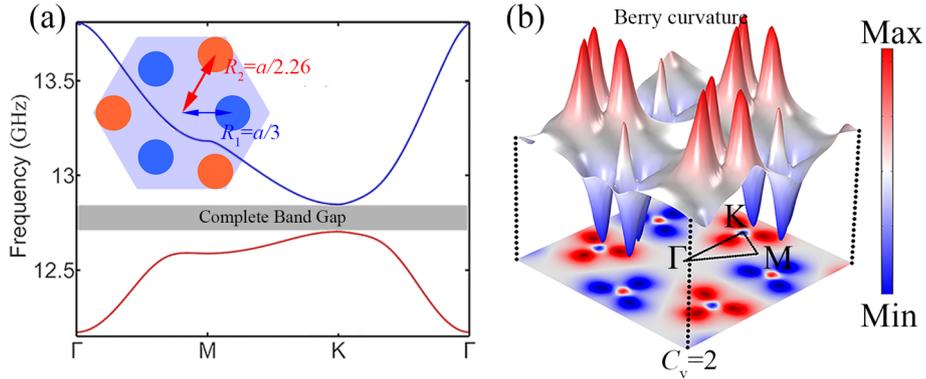

Fig. 2 (a) Band structures of the PC at $R_1 = a/3$ and $R_2 = a/2.26$. (b) The Berry curvature in the first Brillouin zone. The curvature is opposite around the *K* and *K'* points, and the valley Chern number $C_v = C_k - C_{k'} = 1-(-1) = 2$.

A further demonstration of the system having larger valley Chern number is to check the number of topological edge states at the boundary. According to the bulk-edge correspondence [33, 34], the number of edge states between two topologically distinct domains should be the difference of the valley Chern number across the boundary. We construct the domain wall that two domains have opposite valley Chern number $C_v=2$ ($C_k=1$ and $C_{k'}=-1$) and $C_v=-2$ ($C_k=-1$ and $C_{k'}=+1$). The two domains have identical parameters as in Fig. 2(a), but one domain takes a rotation of 60° for its unit cells. Across the domain wall, the differences of the valley Chern number should be $|\Delta C_v|=2$ at *K* (*K'*) point. Indeed, two edge states emerge inside the bandgap as shown in Fig. 3(a). The insets display the edge modes at the dots of the edge dispersion curves. The electric field is localized near the domain wall and decay rapidly away from the wall. As a representative example, figure 3(b) displays the transmission spectra of the Z-shape bend. The waves along the domain wall are against the disorders as shown in the insets of Fig. 3(b). The figure shows that the valley edge states can guide the waves around sharp bent smoothly without reflection in the single-mode region, but with some reflection in the multimode region because the intervalley scattering is increased. In general, simulations agree with our analytical prediction, which further demonstrates the topological phase is with a large valley Chern number of two.

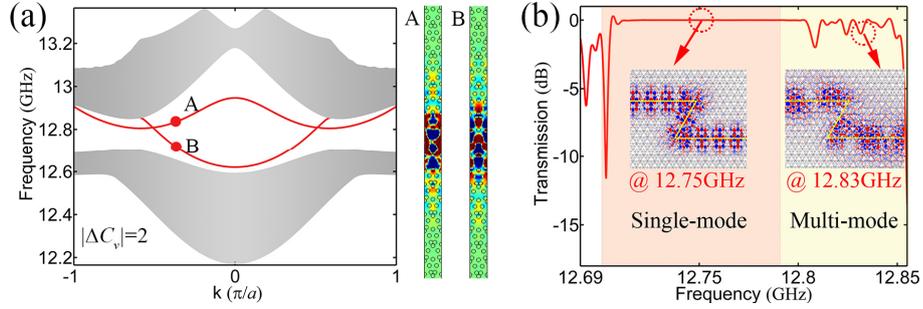

Fig. 3 The topological edge state of valley Chern number $C_v$=2. (a) The projected band structures for the valley Chern difference $|\Delta C_v|$=2 at $K$ ($K$') point across the domain wall. The insets are the distributions of $E_z$ at the given points A and B in the band structure. (b) The transmission spectra of Z-shape corners in the frequency range of 12.69 ~ 12.85 GHz. The red and yellow regions correspond to the single-mode and multi-mode regions, respectively. The inserts are the $E_z$ field distributions at the frequencies in the single-mode and multi-mode regions.

The valley Chern number of the PC depends on the expansion or shrinking of one set of rods. As an example, we keep the blue set of rods at $R_1$=$a$/3, while expanding the red set of rods to $R_2$= $a$/2.36 as shown in the inset of Fig. 4(a). For this situation, the bandgap and energy valley appear in the band structure as shown in Fig. 4(a). At the valley, the phase distribution of electric field $E_z$ and the power flow are illustrated in Fig. 4 (b). The figure shows that the power flows is counterclockwise outside the red rods, but is inside the blue rods. The red and blue rods can be considered as two different trimers, the small trimer and the big trimer, which are in inequivalent position of the unit cells forming A and B sub-lattices. The staggered potential $\Delta$ between the sublattices opens a gap at the Dirac points. A further reflection is the Berry curvature as shown in Fig. 4(c), where three peaks are down near the $K$ point, while three peaks up near the $K'$ point. Thus, the valley Chern number is $C_v=C_k-C_{k'}=-3/2-(3/2)=-3$, a negative number.

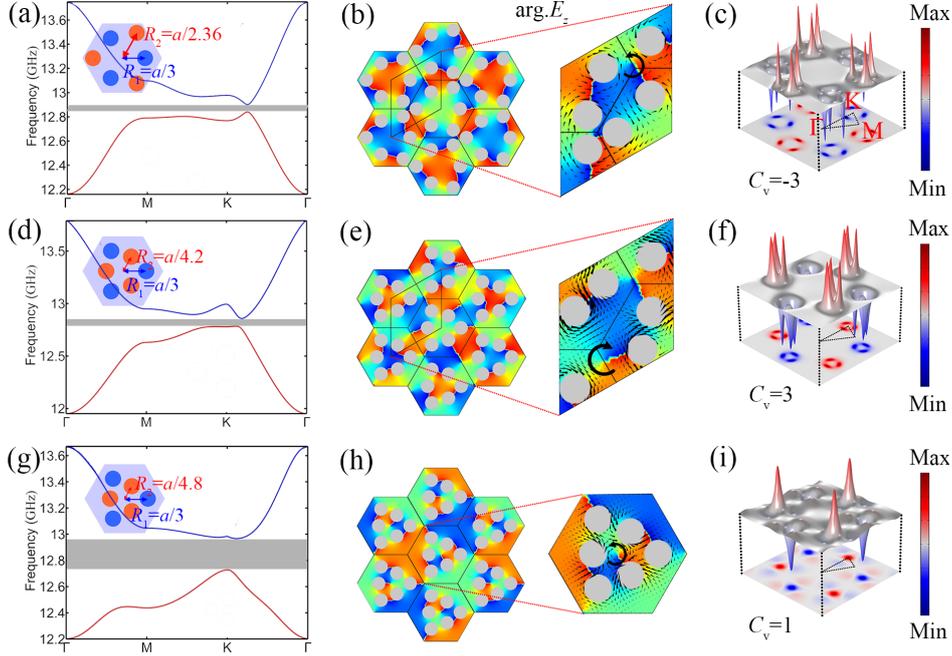

Fig. 4 Band structures, phase and power flow distribution and corresponding Berry curvature of the PC where $R_1$ is fixed at $a/3$ and $R_2$ takes (a) $R_2=a/2.36$ and valley Chern number $C_v=-3$, (b) $R_2=a/4.2$ and $C_v=3$, (c) $R_2=a/4.8$ and $C_v=1$. The arrow inserted in the phase distribution indicates the Poynting vector.

Also, shrinking one subset changes the valley Chern number to a different value. Supposing we reduce the dimension of the red subset to $R_2=a/4.2$. The full bandgap and energy valleys present in the band structure as displayed in Fig. 4(d). The phase distribution and power flows again show the red and the blue trimers can be recognized as two sublattices, as shown in Fig. 4(e). However, the big and small trimers exchange their position in the unit cell that results in opposite staggered potential Δ to the one in Fig. 4(b). This can be observed in the Berry curvature from the Fig. 4(f), where the peaks upward and downward are exchanged between $K$ and $K'$ points to Fig. 4(c). Thus the valley Chern number at $R_2=a/4.2$ is $C_v=C_k-C_{k'}=3/2-(-3/2)=+3$, a positive number.

Further shrinking the dimension of the red set will stronger the coupling between six rods, and all the rods in unit cell function as a whole, the hexamer. As displayed in Fig. 4(h), when $R_2$ takes $a/4.8$ the phase distribution of the electric $E_z$ concentrates at the hexamer and the power flow illustrates a positive OAM locate around the central point of the unit cell, which is similar to the valley phase in the Ref. [27]. The Berry curvature of the band below the bandgap shows only one peak at the $K(K')$ points, and the valley Chern number is $C_v=1$.

The valley Chern numbers are also proved by the number of the topological edge states at the domain wall. Again, we make the domain wall between two PCs with opposite valley Chern number. Fig. 5(a) shows the band structure of the two domains with valley Chern number of $C_v=1$ ($C_k=1/2$ and $C_k=-1/2$) and $C_v=-1$ ($C_k=-1/2$ and $C_k=+1/2$). The two domains have identical geometrical parameters as in Fig. 4(g), but one domain shrinks the red set while the others shrinks blue set. We observe one edge state appearing in the bandgap, because the differences of the valley Chern number $|\Delta C_v|$ across the domain wall is one at $K$ ($K'$) point. Fig. 5(c) illustrates the wave propagates along a Z-shaped domain wall in the frequency range of 12.7 ~ 12.95 GHz. At the edge modes, the wave is localized at the domain wall and robust against the sharp corners of the domain wall. Similarly, for the domain wall created by the PCs of valley Chern number $C_v=-3$ and $+3$, where structure parameters are $R_1=a/3$, $R_2=a/4.2$ for one domain and $R_1=a/4.2$ and $R_2=a/3$ for the other domain, three edge states present in the projected band structure, as shown in Fig. 5(b). The numbers of edge states are consistent with the differences of the valley Chern number across the domain wall; $|\Delta C_v|=3$ at $K$ ($K'$) point. For these edge states, figure 5(b) plots the field profile of the modes, and the field always concentrates on the domain wall. The robustness of the wave propagation along the domain wall in the frequency range of 12.78 ~ 12.87 GHz is shown in Fig. 5(d). We see the wave is efficient going across sharp corners of Z-shaped the domain wall. We should note because of the multiple edge modes at a given frequency, the robustness of the wave propagation may be weaker in some cases of disorders.

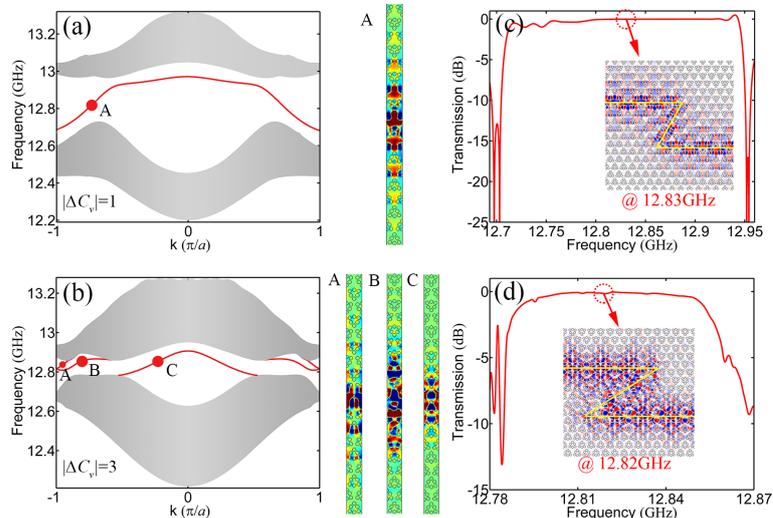

Fig 5 The topological edge state of valley Chern number of $C_v$=1 and 3. Panels (a) and (b) are the projected band structures for the valley Chern difference across the domain wall, respectively, $|\Delta C_v|$=1 and 3 at $K$ ($K'$) point. The number of edge states present in the bandgap are the same as the valley Chern difference. The insets in each panel are the distributions of $E_z$ at the given points in the band structure. The fields are all localized at the domain wall. (c) and (d) The transmission spectra of Z-shape topological domain wall for valley Chern difference $|\Delta C_v|$=1 and 3 at $K$ ($K'$) point, respectively. Simulated $E_z$ field distributions are inserted. The yellow curves represent the energy flows.

In conclusion, we have proposed and theoretically demonstrated a new type of valley Hall phase in 2D photonic crystal made of the hexamers of dielectric rods, which has a large valley Chern number. By simple shrinking or expanding one set of rods in the hexamer, we realize the valley phase transition from the valley Chern number of one to three. The multiple edge states further demonstrate our valley phases having large valley Chern numbers, which are perfectly compatible with the bulk-edge correspondence. Robustness of the edge modes is demonstrated by the wave transmission along the domain wall of the Z-shaped form. Our study provides new opportunities in topological photonics according to the practical requirements.

Acknowledgements: This work is supported by National Natural Science Foundation of China (NSFC) (61771237), and partially by the Project funded by the Priority Academic Program Development of Jiangsu Higher Education Institutions and Jiangsu Provincial Key Laboratory of Advanced Manipulating Technique of Electromagnetic Waves.